\documentclass[twocolumn,epsf,psfig,superscriptaddress]{revtex4}
\usepackage{amssymb}
\usepackage{epsfig}
\DeclareMathAlphabet{\pazocal}{OMS}{zplm}{m}{n}            
\usepackage{graphicx}
\usepackage{color}
\usepackage{bm}
\usepackage[normalem]{ulem}     
\usepackage{soul}
\usepackage{amsmath}
\usepackage{braket} 
\usepackage{rotating}
\usepackage{multirow} 
\usepackage{hyperref}

\begin{document}

\title{
Magnetic Compton profile in non-magnetic ferroelectrics
}

\author{Sayantika Bhowal} 
\affiliation{Materials Theory, ETH Zurich, Wolfgang-Pauli-Strasse 27, 8093 Zurich, Switzerland} 

\author{Stephen P. Collins}
\affiliation{Diamond Light Source Ltd, Diamond House, Harwell Science \& Innovation Campus, Didcot, Oxfordshire, OX11 0DE}

\author{Nicola A. Spaldin}
\affiliation{Materials Theory, ETH Zurich, Wolfgang-Pauli-Strasse 27, 8093 Zurich, Switzerland}

\date{\today}

\begin{abstract}
Magnetic Compton scattering is an established tool for probing magnetism in ferromagnetic or ferrimagnetic materials with a net spin polarization. Here we show that, 
counterintuitively,
{\it non-magnetic} systems can also have a non-zero magnetic Compton profile, provided that space-inversion symmetry is broken. The magnetic Compton profile is antisymmetric in momentum and, if the inversion symmetry is broken by an electric-field switchable ferroelectric distortion, can be reversed using an electric field.  We show that the underlying physics of the magnetic Compton profile and its electrical control are conveniently described in terms of $k$-space magnetoelectric multipoles, which are reciprocal to the real-space charge dipoles associated with the broken inversion symmetry. Using the prototypical ferroelectric lead titanate, PbTiO$_3$, as an example, we show that the ferroelectric polarization introduces a spin asymmetry in momentum space that corresponds to a pure $k$-space magnetoelectric toroidal moment. This in turn manifests in an antisymmetric magnetic Compton profile which can be reversed using an electric field. Our work 
suggests
an experimental route to directly measuring and tuning hidden $k$-space magnetoelectric multipoles via their magnetic Compton profile.    
\end{abstract}

\maketitle

The inelastic Compton scattering of x-ray photons by electrons was an early confirmation of quantum mechanical behavior, in its revolutionary step of assigning momentum to electromagnetic waves \cite{Compton1923}. The effect is widely used today in fields as diverse as radio-biology \cite{Redler2018},  astrophysics \cite{Wiley1985}, and condensed matter, where it is used to measure the distribution of electron density in momentum space \cite{Cooper1971}. The extension to measuring the {\it spin-dependent} momentum distribution, now known as magnetic Compton scattering, was proposed as early as 1970 \cite{PlatzmanTzoar1970}, and dramatic advances in synchrotron light sources mean that the weak scattering cross section in the spin channel of materials can now be captured.  
 
 Over the past decades, magnetic Compton scattering \cite{Cooper2007,Duffy2013,Ahuja2013} has been used effectively to investigate the magnetic properties of ferro/ferrimagnetic materials. The measured quantity, the magnetic Compton profile (MCP), $J_{mag} (p_z)$, is the one-dimensional projection of the spin-polarized electron momentum density 
 which, for backscattering, lies
 along the direction of the scattering vector $p_z$, 
\begin{equation}
J_{mag} (p_z) = \int \int [\rho^\uparrow (\vec p)-\rho^\downarrow (\vec p)] dp_x dp_y \quad.
\end{equation} 
  Here $\rho^\uparrow (\vec p)$  $[\rho^\downarrow (\vec p)]$ is the momentum density of the majority [minority] spin bands. The MCP is insensitive to orbital magnetism due to the instantaneous interaction of the photon with the electrons \cite{Timms1993, Cooper1992}, and therefore allows separation of the spin and orbital contributions to the total magnetic moment \cite{Cooper1991,Duffy2010,Itou2013}, as well as the magnetic contributions from localized and itinerant electrons \cite{Zukowski1993,Duffy1998,Duffy2000,Banfield2005,ShentonTaylor2007}. It is also used to measure the spin polarization at the Fermi surface \cite{Duffy2013,Mijnarends2007,Mizoroki2011}, which is at the heart of spintronic applications. Since a non-zero MCP
  was
  believed to require broken time-reversal ($\cal T$) symmetry, it has to date only been studied in magnetic materials. 
 
  Here we show that a non-zero MCP can exist in non-magnetic (NM) materials provided that space-inversion ($\cal I$) symmetry 
  is broken.  The MCP in this case is antisymmetric, since the $\cal T$ symmetry dictates $J_{mag} (p_z)\overset{\cal T}{\rightarrow} -J_{mag} (-p_z)$. This behavior is in striking contrast to the centrosymmetric, NM case with both  $\cal I$ and $\cal T$ symmetries, for which the MCP is zero since $J_{mag} (p_z)\overset{\cal I}{\rightarrow} J_{mag} (-p_z)$.

\begin{figure}[t]
\centering
%\begin{flushleft}
\includegraphics[width=\columnwidth]{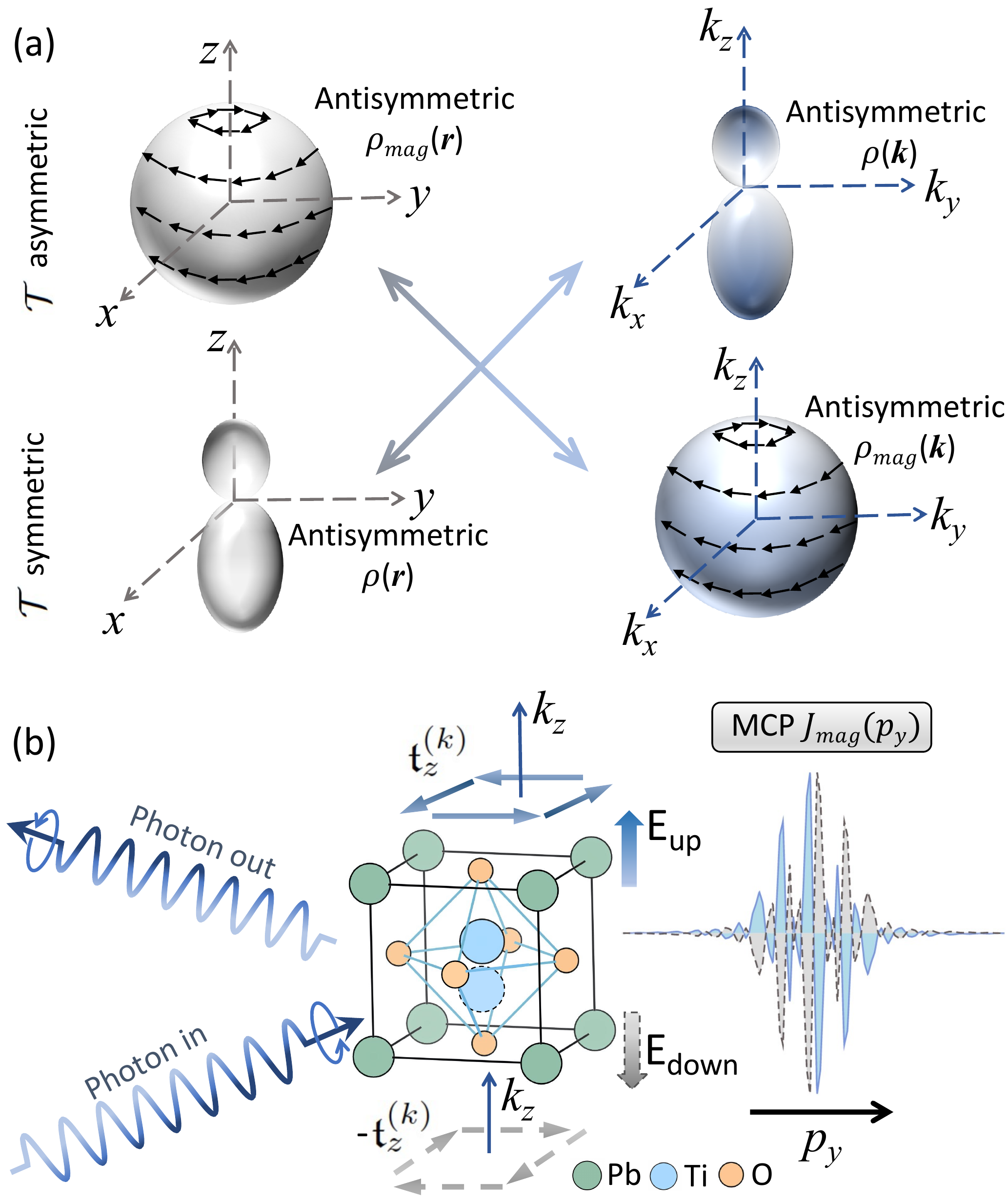}
%\includegraphics[scale=.5]{fig1}
% \end{flushleft}
 \caption{(a) Cross-links between real ({\it left}) and $k$-space ({\it right}) multipoles in $\cal T$ asymmetric ({\it top}) and symmetric ({\it bottom}) odd-parity systems. (b) Magnetic Compton scattering with a circularly polarized photon beam in PTO. The solid-outlined Ti-atom position, $\mathfrak{t}^{(k)}_z$ spins and MCP profile indicate one polarization orientation. With reversed E (dashed outlines), the Ti ion displacement reverses, the toroidal moment $\mathfrak{t}^{(k)}_z \rightarrow -\mathfrak{t}^{(k)}_z$, and the sign of the MCP reverses. }
 \label{fig1}
 \end{figure}

These simple symmetry relations indicate the possibility of a non-zero MCP for a NM system, and the underlying physics can be elegantly described in terms of the odd-parity electric multipoles induced by the broken $\cal I$ symmetry, which determine both the occurrence of a non-zero $J_{mag} $ and its direction in momentum space. 
The key physics is illustrated in Fig. \ref{fig1}, showing the duality between real and $k$ space \cite{Watanabe2018,BhowalSpaldin2021} in odd parity ($\cal I$ asymmetric) materials, between the real- and momentum-space multipole representations in magnetic and NM species. As seen from the top panel of Fig. \ref{fig1} (a), the real-space ME multipole (MEM) (antisymmetric under both $\cal I$ and $\cal T$) \cite{ClaudeSpaldin2007,Spaldin2008,Spaldin2013} corresponds to an electric dipole (no magnetization dependence) in reciprocal space, which gives rise to an antisymmetric part in the {\it regular} Compton profile \cite{BhowalSpaldin2021}. Here we point out that an exactly opposite situation exists for a real-space electric dipole (bottom panel of Fig. \ref{fig1} (a)), which corresponds to a MEM in momentum space and, in turn, gives rise to an antisymmetric MCP.

The transformation of the electric dipole in real space into a MEM in $k$-space follows from the fact that both $r$ and $k$ change sign upon inversion,  i.e.\  $\vec{r} \overset{\cal I}{\rightarrow} -\vec{r} $, and $\vec{k} \overset{\cal I}{\rightarrow} -\vec{k} $, but behave differently under time-reversal, with $\vec{r} \overset{\cal T}{\rightarrow} \vec{r} $, but $\vec{k} \overset{\cal T}{\rightarrow} -\vec{k} $. As a result, broken inversion symmetry in real space alone is sufficient to break both inversion and time-reversal symmetries in momentum space, leading to $k$-space MEMs.

The following implications of these $k$-space MEMs are key to our work: First, since the odd-parity charge multipoles [bottom panel of Fig. \ref{fig1} (a)], which measure the asymmetry of the electron density in real space ({\it left}), create an asymmetry in the magnetization density in $k$ space ({\it right}), they can be probed using magnetic measurements in momentum space, such as magnetic Compton scattering. Second, as the $k$-space MEMs are the measure of this momentum-space asymmetry in the magnetization density, the MCP provides a direct signature of the specific $k$-space MEMs. 
Third, since the $k$-space MEMs originate from broken $\cal I$ symmetry, structural tuning that changes these multipoles will modify the MCP. Such structural changes can be easily accessible in ferroelectric materials, where the lattice distortions (electric polarization) can be controlled by an external electric field. Such an electrical tuning of the MCP typically allows faster switching than a conventional magnetic field, leading to reduced systematic errors in experiments.

Here, we illustrate the above ideas by explicitly computing the MCP and the odd parity multipoles for the prototype tetragonal ferroelectric PbTiO$_3$ (PTO), using first-principles methods based on density functional theory (DFT) as implemented in extended versions \cite{Ernsting2014,Spaldin2013} of the Elk code \cite{code}. The odd-parity charge multipoles are extracted by decomposing the $\cal T$ symmetric density matrix $\rho_{lm,l'm'}$ into parity-odd tensor moments, where only the odd $l-l'$ terms contribute to the desired multipoles  \cite{Spaldin2013}. To illustrate how different multipoles result in a different MCP, we also introduce the case of the rhombohedral ferroelectric GeTe.   

The key finding of our work is the antisymmetric MCP for NM ferroelectrics, and its switching using an applied electric field (Fig. \ref{fig1} (b)). We show that odd-parity charge multipoles and spin-orbit interaction (SOI) are the key ingredients for MCP in broken $\cal I$-symmetry systems. We reveal how the structural asymmetry, governed by the odd-parity charge multipoles, introduces additional inter-orbital hopping parameters. These cause the magnetic-moment dependence of the charge multipoles in $k$ space, and the MEMs in momentum space. An interesting outcome is the realization of a pure ME toroidal moment, $\mathfrak{t}^{(k)}_z$, in the momentum space of PTO; such pure toroidal moments are hard to realize in real space, where they are usually accompanied by ME quadrupole moments \cite{Spaldin2008,Spaldin2013,Florian2020,Spaldin2021,BhowalSpaldin2021}.  The magnetic asymmetry, governed by these $k$-space MEMs, combined with SOI, results in the antisymmetric MCP. 

We start by analyzing the symmetry of ferroelectric PTO, which crystallizes in the tetragonal $P4mm$ structure \cite{Nelmes1985}, with no ${\cal I}$ symmetry [middle panel of Fig. \ref{fig1} (b)]. The absence of ${\cal I}$ in the $C_{4v}$ point group allows for several odd-parity charge multipoles, of which the $A_1^+$ irreducible representation (IR), that corresponds to the PTO structure, allows for the odd parity $Q_{10}$ dipole as its lowest order charge multipole (note that higher order multipoles of odd parity, such as the $Q_{30}$ octupole, etc. are  also allowed) as shown in Table I \cite{note1,Watanabe2018,books,Bilbao}. As expected from the duality and cross-links discussed above, the corresponding $k$-space representation has explicit dependence on the magnetic moment ($\vec \mu$) and represents a pure $k$-space toroidal moment $\mathfrak{t}_z^{(k)} = k_x \mu_y - k_y \mu_x$ (this can be seen simply by replacing $\vec r \rightarrow \vec k$), without any ME quadrupole moment components. In fact, the $k$-space ME quadrupole moments belong to different IRs  (${\cal Q}_{xy}^{(k)}$ to $B_1^{+}$, etc). By definition, the existence of $\mathfrak{t}_z^{(k)}$ indicates the presence of magnetically polarized bands with magnetic orientation along $y$ ($x$) in the $k_x$ ($k_y$) direction of momentum space, which, in turn, implies the presence of a MCP along $p_x$ ($p_y$). Here, $x, y, z$ denote the Cartesian axes. 

 \begin{table} [t]
\caption{The basis functions of the odd parity charge multipoles (lowest order in $k$) corresponding to the IR representations of PTO and GeTe. The $k$-space ME monopole ${\cal A}^{(k)}$, toroidal moment $\vec{\mathfrak{t}}^{(k)}$, and quadrupole moment ${\cal Q}_{ij}^{(k)}$ are also indicated within the parentheses. The magnetic moment $\vec \mu \equiv \mu_B \big(\frac{2\vec l}{l+1} +2\vec s \big)$ includes contributions from both spin $\vec  s$ and orbital $\vec l$ angular momenta.}
\centering
\begin{tabular}{ c c c   c }
\hline
IR    &  Multipole  & Real space   & $k$ space  \\ 
%     &       &        &   \\ [1 ex]
\hline
\multicolumn{4}{c}{PTO ($C_{4v}$)} \\[1 ex]
\hline
$A_1^{+}$ &   $Q_{10}$   &  $z$ &  $k_x\mu_y-k_y\mu_x$ ($\mathfrak{t}_z^{(k)}$)  \\
%    &        &      &       \\
%
\hline
\multicolumn{4}{c}{GeTe ($C_{3v}$)} \\
\hline
 $A_1^{+}$ &   $Q_{10}$   &  $z$ &  $k_x\mu_y-k_y\mu_x$($\mathfrak{t}_z^{(k)}$)   \\
 $E^+$    &   $\{Q_{11}^+, Q_{11}^-\}$      &  $\{ x,y \}$    & $\{k_y \mu_z + k_z \mu_y ,k_z \mu_x + k_x \mu_z\}$       \\ 
%
 %          & & & $k_z \mu_x + k_x \mu_z\} $  \\
           & & &  ($\{ {\cal Q}_{yz}^{(k)}, {\cal Q}_{xz}^{(k)} \}$)\\
           & $\{ Q_{31}^+, Q_{31}^-\} $ & & $\{k_y \mu_z - k_z \mu_y,k_z \mu_x - k_x \mu_z\}  $ \\
 %          &                            & & $k_z \mu_x - k_x \mu_z\}$ \\
            &                           &  & ($\{ \mathfrak{t}_x^{(k)}, \mathfrak{t}_y^{(k)}\}$)              \\
             & $\{ Q_{32}^+, Q_{32}^-\} $ & $\{ (x^2-y^2)z,$ & $\{k_x\mu_y+k_y\mu_x, k_x\mu_x -k_y\mu_y \}$   \\
%             &       & $xyz \}$                     & $k_x\mu_x -k_y\mu_y \} $ \\
             &       &    $xyz \}$                     &  $(\{ {\cal Q}_{xy}^{(k)}, {\cal Q}_{x^2-y^2}^{(k)} \})$     \\
  \hline
\end{tabular}
\label{tab1}
\end{table}

\begin{figure}[t]
\centering
%\begin{flushleft}
\includegraphics[width=\columnwidth]{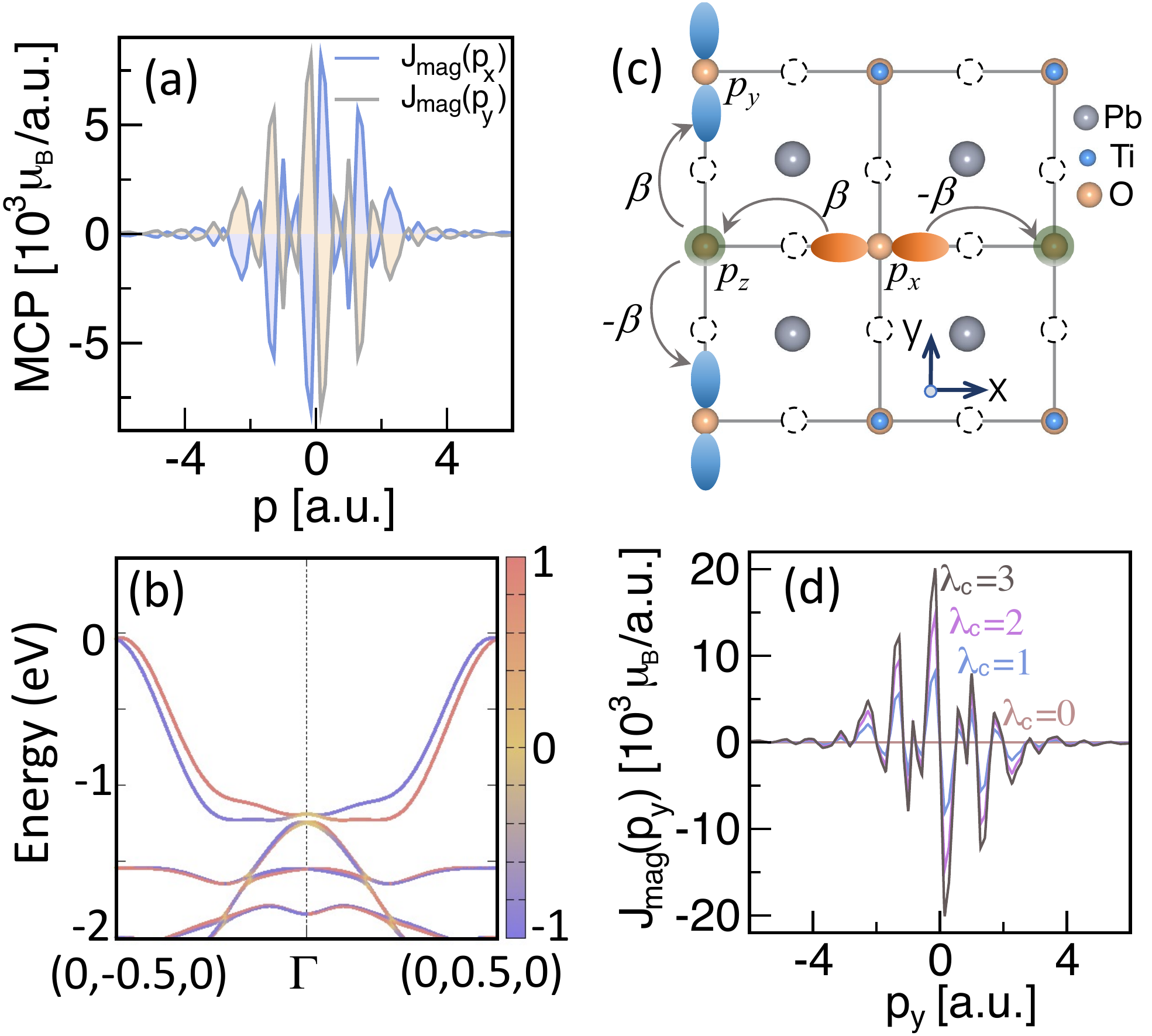}
%\includegraphics[scale=.5]{fig1}
% \end{flushleft}
 \caption{(a) Calculated MCPs of PTO along the $p_x$ and $p_y$ directions in momentum space. (b) Bandstructure of PTO, showing the same spin asymmetry in the $S_x$ spin component, shown in color map, along $\pm k_y$. The Fermi energy is set at zero eV. (c) Illustration of the antisymmetric hopping $\pm \beta$ between $p_x$-$p_z$ and $p_y$-$p_z$ orbitals of the O atoms along $\pm \hat x$ and $\pm \hat y$ directions respectively, induced by broken $\cal I$ symmetry. Only hoppings between apical O atoms are indicated. In-plane O atoms are shown in dashed circles. (d) Dependence of the MCP on the enforced SOI strength $\lambda = \lambda_c \times \lambda_r$, with $\lambda_r$ and $\lambda_c$ the actual SOI strength and the scaling factor respectively.}
 \label{fig2}
 \end{figure}

Next, we explicitly compute the MCP of PTO \cite{ComputaionalDetails}
and show our result in Fig. \ref{fig2} (a). We note first that the computed MCPs satisfy  $m=\int_{-\infty}^\infty J_{mag} (p_x) dp_x = \int_{-\infty}^\infty J_{mag} (p_y) dp_y = 0$, as required for a NM system, since they are purely antisymmetric functions. The computed MCP is about three orders of magnitude smaller than the corresponding total Compton profile and is about an order of magnitude smaller than the MCP for ferromagnetic Ni \cite{Dixon1998}. The signs of the MCPs along $p_x$ and $p_y$ are opposite, directly reflecting the opposite signs of $k_x$ and $k_y$ in $\mathfrak{t}_z^{(k)} = k_x \mu_y - k_y \mu_x$. 
These asymmetries are also reflected in the spin asymmetry in the calculated bandstructure along the $\pm k_x$ (or $\pm k_y$) direction of the momentum space, shown in Fig. \ref{fig2} (b). This spin asymmetry is also consistent with the presence of $\cal T$ symmetry, which dictates that the $\uparrow$-spin band at $+\vec k$ has the same energy as the $\downarrow$-spin band at $-\vec k$, that is $\varepsilon_n (\vec k \uparrow) = \varepsilon_n (-\vec k \downarrow)$, creating the left-right asymmetry of the spin-polarized bands in Fig. \ref{fig2} (b). 

 Since the spin asymmetry that leads to the MCP is largest at the top of the valence bands, which are predominantly of O-$p$ character,  we next analyze the role of the O $p$ orbitals in the chemical bonding, by analyzing the hopping parameters between neighboring O $p$ orbitals along $\hat x$ and $\hat y$ directions. 
We note that the off-centering of the atoms in the non-centrosymmetric ferroelectric structure induces additional interorbital hoppings between the $p_x$-$p_z$ and $p_y$-$p_z$ orbitals of the O atoms, which are forbidden in the Slater-Koster \cite{SlaterKoster1954,Harrison1989} parameters in the presence of $\cal I$ symmetry [see Fig. \ref{fig2} (c)]. Interestingly, the computed effective hopping parameters for the broken symmetry structure, extracted using the NMTO downfolding method \cite{AndersenSaha-Dasgupta}, have left-right asymmetry (Fig. \ref{fig2} (c)). Such antisymmetric hopping parameters lead to sine terms in the tight-binding (TB) model, in contrast to the usual cosine terms associated with symmetric hopping, e.g. $\langle p_x |{\cal H_A}|p_z \rangle = -2i\beta \sin (k_xa)$. In the small $k$ limit, $\sin(k_xa) \rightarrow k_xa$, and the corresponding $p$-orbital TB Hamiltonian is ${\cal H_A} (k) =\gamma (k_x\hat L_y-k_y\hat L_x)$, where the constant $\gamma \propto \beta$ is a measure of $\cal I$ asymmetry and $\hat L_x, \hat L_y$ are the Cartesian components of the orbital angular momentum operator for the $p$ orbitals \cite{SuganoTanabeKamimura1970}. 
We recognize that ${\cal H_A}$ represents the orbital part of the toroidal moment $\mathfrak{t}^{(k)}_z$. In the presence of SOI, the orbitals couple with the spins, giving rise to a spin component of $\mathfrak{t}^{(k)}_z$.    
Since the MCP is insensitive to orbital magnetism, we see here the importance of SOI in the MCP of PTO. To further clarify the role of SOI in MCP, we next artificially change the strength of the coupling $\lambda$ in our calculations. As seen from Fig. \ref{fig2} (d), this results in a drastic change in the magnitude of the MCP, with vanishing MCP at $\lambda=0$, consistent with the hopping analysis. 

\begin{figure}[t]
\centering
%\begin{flushleft}
\includegraphics[width=\columnwidth]{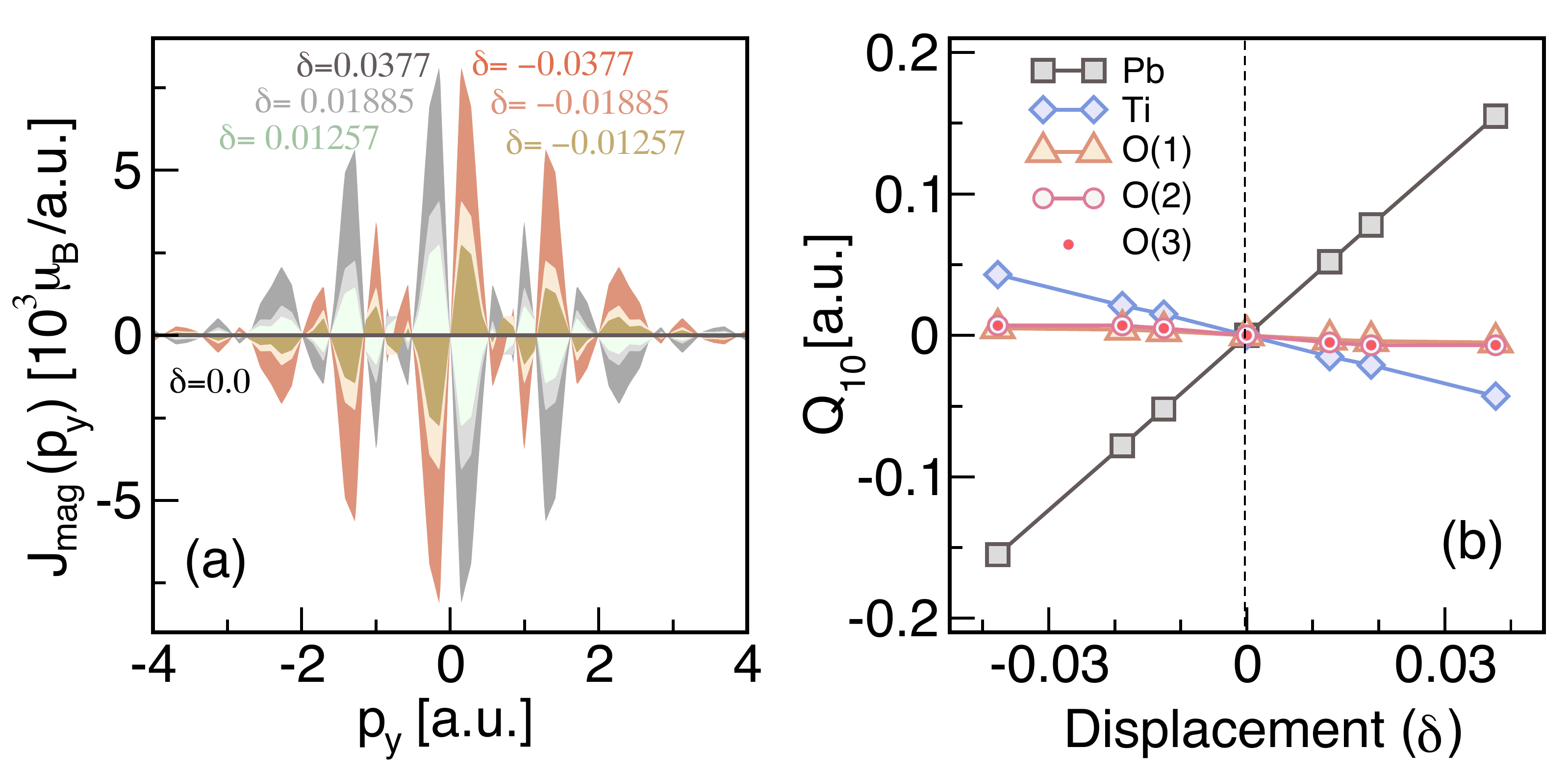}
%\includegraphics[scale=.5]{fig1}
% \end{flushleft}
 \caption{(a) Variation of MCP with displacement $\delta$ (in units of lattice vector $c$) of the Ti atom from the centrosymmetric position in PTO. The MCP increases with $\delta$, and switches sign as the sign of $\delta$ changes. (b) Variation of the odd-parity charge multipole $Q_{10}$ at Pb, Ti and O sites [apical O(1) and in-plane O(2) and O(3)] for the same displacements $\delta$ as in (a). The correlation between $Q_{10}$ and MCP is apparent. The vertical dashed line corresponds to the centrosymmetric case with  $\delta=0$, at which both multipoles and MCP vanish.
 The odd-parity multipoles at Pb and Ti sites are obtained by adding the $p$-$d$ and $p$-$s$ contributions, while for the O atoms only the $p$-$s$ contribution is considered.}
 \label{fig3}
 \end{figure}

We now turn to the tuning of the MCP by manipulating the odd-parity MEMs through changing the charge multipoles. For this, we artificially change the ferroelectric displacements of the atoms in PTO along the polarization direction $[001]$. The resulting changes in MCP are depicted in Fig. \ref{fig3} (a). We also study the corresponding variation in the charge multipoles at Pb, Ti and the three O atoms, shown in Fig. \ref{fig3} (b). As expected, the multipole at the Pb atom dominates due to the $6s^2$ lone pair that causes the broken $\cal I$ symmetry. The magnitudes of the multipoles at the Ti and O atoms are rather weak and have opposite sign to that of Pb atom. 
 It is also apparent from Figs. \ref{fig3} (a) and (b) that the MCP follows the trend in odd parity multipoles: The magnitudes of both increase with increasing ferroelectric displacement and both reverse sign as the displacement is switched. Since the switching of ferroelectric displacements in PTO can be achieved using an external electric field, our results show the possibility of switching MCP using electrical means, which may be of practical importance in reducing the experimental uncertainties in a measurement of the effect. Reciprocally, the MCP provides a possible probe for the detection of odd parity electric multipoles.

\begin{figure}[t]
\centering
%\begin{flushleft}
\includegraphics[width=\columnwidth]{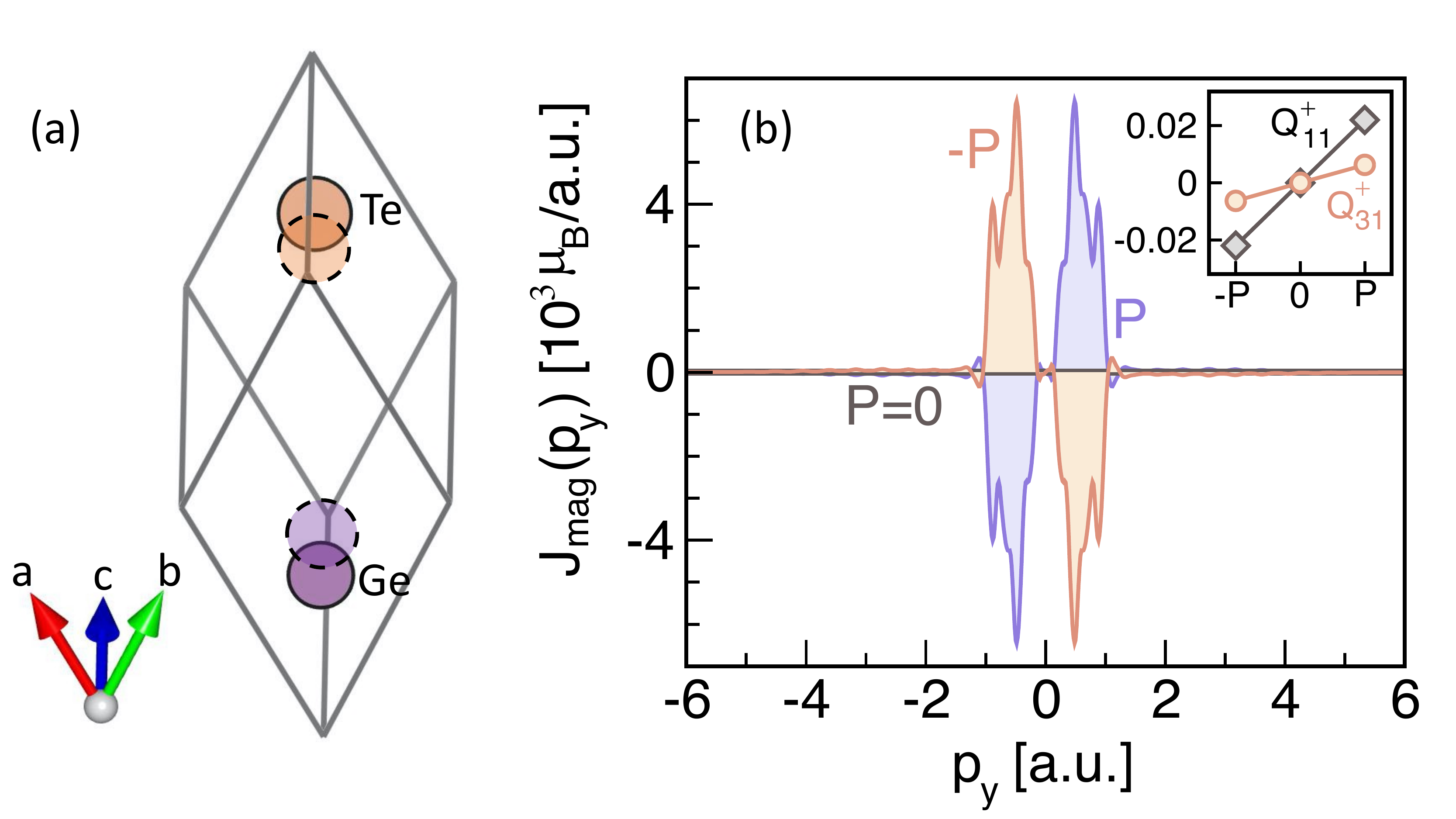}
%\includegraphics[scale=.5]{fig1}
% \end{flushleft}
 \caption{Switching of MCP in GeTe. (a) The rhombohedral unit cell of GeTe. The dashed atoms indicate the centrosymmetric Ge (0.25,0.25,0.25) and Te (0.75,0.75,0.75) positions; the solid atoms the positions with polarization P. (b) Calculated MCP for structures with polarization P, -P and 0 (for which the MCP vanishes). Inset: odd-parity multipoles Q$_{11}^+$ and Q$_{31}^+$ (in a.u.), which contribute to $J_{mag} (p_y)$, for the antisymmetric spin density with spin quantization along [001]. Multipoles at Ge and Te have opposite sign. Q$_{11}^+$ is summed from its $p$-$d$ and $p$-$s$ contributions; Q$_{31}^+$ has only $p$-$d$ contribution.}
 \label{fig4}
 \end{figure}

To emphasize the correlation between MCP and odd-parity multipoles, we now briefly analyze ferroelectric GeTe, with structural polarization along the [111] direction of the rhombohedral unit cell (Fig. \ref{fig4} (a)). The ferroelectric distortion of GeTe corresponds to $A_1$ and $E$ IRs of its $C_{3v}$ point group symmetry \cite{Chatterji2018,Kagdada2018}. These allow the $\mathfrak{t}_x^{(k)}, \mathfrak{t}_y^{(k)}$ components of the $k$-space toroidal moment, as well as quadrupole moment components ${\cal Q}_{xy}^{(k)},{\cal Q}_{yz}^{(k)},{\cal Q}_{xz}^{(k)},$ and ${\cal Q}_{x^2-y^2}^{(k)},$ in addition to $\mathfrak{t}_z^{(k)}$ [see Table \ref{tab1}], and the corresponding MCP 
has more components 
than in the case of PTO. First, in contrast to PTO, the MCPs in GeTe have simultaneous contributions from toroidal and quadrupole moments rather than pure toroidal contribution. For example, $\{ \mathfrak{t}_x^{(k)}, {\cal Q}_{yz}^{(k)} \}$ contribute to the MCP along $p_y$ with [001] spin quantization axis. 
Secondly, even along the same momentum direction, the MCP can appear for different spin quantization axes. For example, along the $p_y$ direction in momentum space, an MCP  exists for $\hat x, \hat y,$ and $\hat z$ spin quantization axes. Knowledge of the odd-parity multipoles is invaluable in interpreting such a complex MCP. We see that the MCP can again be switched by reversing the direction of polarization [see Fig. \ref{fig4} (b)]. While the magnitude of the MCP is slightly smaller than that of PTO, in GeTe MCP components 
with parallel spin and momentum directions [e.g., along $\hat x$ and $\hat y$ driven by ${\cal Q}_{x^2-y^2}$ (see Table \ref{tab1})] exist, allowing for experimental detection  within the simple back-scattering geometry.  

Finally, we discuss the experimental set up for detecting the MCP in PTO. The experiment will be similar to conventional magnetic Compton scattering, with a circularly polarized beam \cite{Cooper2007}, except that no magnetic field is required. The antisymmetric spin profile can be obtained by reversing
the circular polarization of the incoming photon and subtracting the two signals, similar to the case of antisymmetric Compton measurements \cite{Bhowal2021}, or, 
more conveniently, the electric polarization can
be flipped rapidly with fixed photon helicity by reversing the
electric field. 
Sensitivity to the MCP arises from a relativistic correction to the scattering cross section (details are given in Ref. \cite{Cooper2007}). For the ideal case of backscattering, the relative sensitivity to the MCP, compared to to the NM profile, is reduced by $\sim \frac{2E}{m_ec^2}$, where $E$ is the incident photon energy. For 100 keV photons, this amounts to a relatively modest reduction of $\sim$0.4.

It is important to note that we need a single ferroelectric domain of PTO or GeTe for the magnetic Compton scattering measurements, which may be obtained at room temperature (the ferroelectric transition temperatures for PTO and GeTe being T$_{\rm c} \sim 765$ K \cite{Samara1971} and around 650 K\cite{Chattopadhyay1987,Wdowik2014} respectively) using an electric field. The switching of the MCP can be observed  by reversing the direction of the electric field, which switches the ferroelectric domain, and, hence, the odd parity multipoles.  
   
The odd parity real-space charge multipoles and associated $k$-space MEMs also have implications in other physical effects beyond the MCP \cite{Hayami,Watanabe2020,Onimaru,Fu,Hayami2020}. Examples include an exotic superconducting state driven by odd-parity multipole fluctuations \cite{Ishizuka2018,Sumita2020}, and the predicted Rashba effect in PTO \cite{Arras2019}, which is a direct consequence of $\mathfrak{t}_z^{(k)}$.  Higher order multipoles in PTO can  give rise to a Berry curvature dipole which should lead to a nonlinear Hall effect. Furthermore, the $k$-space MEMs could describe the recently observed polar skyrmions \cite{Das2019}, in the same spirit as their real-space counterpart forms a basis for describing magnetic skyrmions \cite{Ingrid2019}.  We hope that our proposal stimulates experimental efforts to measure the MCP, or these related MEM-driven  behaviors, in NM ferroelectrics.

\section*{Acknowledgements}
We thank Jon Duffy, Stephen Dugdale and Urs Staub for stimulating discussions.  
NAS and SB were supported by the ERC under the EU’s Horizon 2020 research and innovation programme grant No 810451 and by the ETH Zurich. Computational resources were provided by ETH Zurich's Euler cluster, and the Swiss National Supercomputing Centre, project ID eth3.

\bibliographystyle{apsrev4-1}
\bibliography{LNPO}

 %\newpage

\end{document}